\documentclass[conference]{IEEEtran}
\IEEEoverridecommandlockouts
\usepackage{cite}
\usepackage{amsmath,amssymb,amsfonts}
\usepackage{algorithmic}
\usepackage{graphicx}
\usepackage{textcomp}
\usepackage{xcolor}
\usepackage{graphicx}
\usepackage{float}
\usepackage{subfig}
\def\BibTeX{{\rm B\kern-.05em{\sc i\kern-.025em b}\kern-.08em
    T\kern-.1667em\lower.7ex\hbox{E}\kern-.125emX}}

\begin{document}

\title{Information Importance-Aware Defense against Adversarial Attack for Automatic Modulation Classification:An XAI-Based Approach
\thanks{This work was supported in part by the National Science Foundation of
	China under Grants 62231015 and 62101253, the Natural Science Foundation
	of Jiangsu Province under Grant BK20210283, and the open research fund of National Mobile Communications Research Laboratory, Southeast University (No. 2022D08). (Corresponding author: Peihao Dong.)}
}

\author{\IEEEauthorblockN{Jingchun Wang$^*$, Peihao Dong$^*$$^\dagger$, Fuhui Zhou$^*$, Qihui Wu$^*$}
	\IEEEauthorblockA{$^*$ College of Electronic and Information Engineering, Nanjing University of Aeronautics and Astronautics, Nanjing 211106, China\\
		$^\dagger$ National Mobile Communications Research Laboratory, Southeast University, Nanjing 211111, China\\
		Email: \{wjc797952, phdong\}@nuaa.edu.cn, zhoufuhui@ieee.org, wuqihui2014@sina.com}
}

\maketitle

\begin{abstract}
%In the realm of wireless communications, fast and accurate modulation identification is crucial. However, traditional approaches face challenges in terms of non-ideal conditions and high complexity. The advent of 
Deep learning (DL) has significantly improved automatic modulation classification (AMC) by leveraging neural networks as the feature extractor.
However, as the DL-based AMC becomes increasingly widespread, it is faced with the severe secure issue from various adversarial attacks. 
Existing defense methods often suffer from the high computational cost, intractable parameter tuning, and insufficient robustness.
This paper proposes an eXplainable artificial intelligence (XAI) defense approach, which uncovers the negative information caused by the adversarial attack through measuring the importance of input features based on the SHapley Additive exPlanations (SHAP).
By properly removing the negative information in adversarial samples and then fine-tuning(FT) the model, the impact of the attacks on the classification result can be mitigated.
Experimental results demonstrate that the proposed SHAP-FT improves the classification performance of the model by 15\%$\sim$20\% under different attack levels,
which not only enhances model robustness against various attack levels but also reduces the resource consumption, validating its effectiveness in safeguarding communication networks.
\end{abstract}


\section{Introduction} 
Traditional modulation identification techniques rely on feature extraction, statistical pattern recognition, and decision theory\cite{b1}. However, these methods may lead to poor accuracy in non-ideal channel conditions or mismatched actual channel characteristics.

The rapid development of deep learning (DL) has significantly advanced the signal automatic modulation classification (AMC)\cite{b2,b3,b4,b5,b6}. DL models surpass the limitations of conventional methods and excel in extracting the hidden features, yielding the high performance in a low-complexity manner.
Although the DL-based solution offers many benefits, it is still vulnerable to various adversarial attacks.
Prior works have demonstrated that imperceptible changes to input data can significantly degrade a model's performance or even completely mislead its decisions\cite{b7,b8}. 
%For instance, a minor pixel perturbation on an image of a truck can cause it to be misclassified as an ostrich\cite{b8}. 
In wireless communications, attackers can exploit such perturbations to alter transmitted message content and further to the misclassification. Consequently, developing effective defense mechanisms against these adversarial attacks is essential for the implementation of DL-based solutions.

Common white-box attack methods can be categorized into the gradient optimization-based attack (e.g., fast gradient sign method (FGSM)\cite{b7} and the projected gradient descent (PGD)\cite{b9}), the decision boundary analysis-based attack (e.g., DeepFool\cite{b10}), and the optimization-based attack (e.g., Carlini-Wagner Attack)\cite{b11}. To counter FGSM attacks, \cite{b7} and \cite{b12} proposed the adversarial training and defensive MGAN using generative adversarial networks. Additionally, \cite{b13} developed a dual defense mechanism to address FGSM attacks. For multi-species attacks, \cite{b14,b15,b16} manipulated the inputs to enhance the model robustness.
Although existing defense methods have enhanced model performance to varying degree. Most of them require substantial computational resources and exhibit the limited generalization ability. Additionally, some approaches struggle with the intractable parameter tuning and insufficient robustness.

These problems can be addressed by defending against network attacks using eXplainable artificial intelligence (XAI) theories and methods. Initial research on XAI focused on the black-box problem in machine learning. XAI reveals the causality of a model through the visualization, logical reasoning, or model transparency, thereby increasing the model's trustworthiness.

Shapley additive explanations (SHAP)\cite{b17} has become the most widely used post-hoc explanation method in XAI due to its standardized and rigorous theory. Based on the game theory of shapley value and additive feature attribution explanation, SHAP calculates the impact of each input feature on the network output. The SHAP method can effectively measure the importance of input features, providing insightful guidance for boosting model performance.
Therefore, the SHAP is promising to improve the DL-AMC by helping understand the enabling model. To our best knowledge, few related works have been reported in the existing literature.

In this paper, we develop an SHAP-FT based defense approach for DL-AMC models deteriorated by the adversarial attack. The main contributions can be summarized as follows:

\begin{itemize}
\item By exploiting SHAP, the importance of the input information can be fundamentally measured, based on which the dominated negative information related to the attack is identified.
\item Through the further analysis and manipulation, the dominated negative information deteriorating the classification accuracy is properly removed, followed by the fast model fine-tuning to retrieve the performance loss.
\item Extensive simulation results based on the widely used dataset are presented to validate the effectiveness of the SHAP. The proposed SHAP-FT improves the model's classification performance by 15\%$\sim$20\% under different attack levels.

%as well as the superiority of the proposed SHAP-FT in defending against the adversarial attack. 
%The results show that the proposed SHAP-FT improves the model's classification performance by 15\% to 20\% under different attack levels.
\end{itemize}

\section{Preliminaries}
\subsection{SHAP}
The SHAP method is based on additive feature attribution interpretation and the shapley value from game theory, allowing it to attribute the model's output to the contribution of each input feature. Due to its theoretical advantages of local accuracy, missingness, and consistency, SHAP has become the most widely used method in XAI.
 Simple models have their own interpreters, while complex deep networks do not. LIME\cite{b19} proposes constructing a simple model $g$ to fit the original complex model $f$, mapping the simple model's input $x'$ to the original input through $x=h_{x}(x')$ with the condition $z\approx x'$ for $g(z)\approx f(h_{x}(z))$.

The output of the simple model can be explained through additive feature attribution, attributing the model's output to each feature's contribution
\begin{equation}
	g(z)=\beta_{0}+\beta_{1}z_{1}+...+\beta_{p}z_{p}, \label{eq}
\end{equation}
where $g(\cdot)$ represents a simple fitting model, $z_{i}$ denotes the $i$-th input feature,  $\beta_{i}$ is the weight contribution of the input feature. Therefore, the contribution increment of the $j$-th feature can be expressed as
\begin{equation}
	\phi_{j}(g)=\beta_{j}Z_{j}-\beta_{j}E(Z_{j}), \label{eq}
\end{equation}
where $E(Z_{j})$ represents the expectation of the feature. Consequently, the total contribution needing allocation is
\begin{equation}
	\begin{split}
		\sum_{j=1}^{p}\phi_{j}(g)&=\sum_{j=1}^{p}(\beta_{j}Z_{j}-\beta_{j}E(Z_{j}))\\
		&=g(z)-E(g(z)).\label{eq}
	\end{split}
\end{equation}
The shapley value was first introduced to address the cooperative game problem in game theory, specifically aimed at achieving a fair distribution of each player's contribution. For a set $N=\{1,2,..., n\}$ of $n$ input units ($n$ players), the shapley value of the $i$-th input unit denotes the $i$-th player's contribution. The calculation formula is as follows
\begin{equation}
	\phi_{i}=\sum_{S\subseteq N{i}}\frac{(|N|-|S|-1)!|S|!}{|N|!}[v(S)-v(S\backslash\{i\})], \label{eq}
\end{equation}
where, $S$ represents different sets of input units, with $S$ being a subset of $N$. $v(S)$ indicates the output value of the network when the input units in $S$ are utilized. For the $i$-th input unit,$v(S\backslash\{i\})$ denotes the output value after removing the $i$-th input unit from set $S$, illustrating the impact of excluding that specific input unit. The right term in the formula portrays the marginal contribution of the $i$-th input unit within the set $S$. 
On the left side of the subtraction term, the coefficient signifies the ratio of random permutations of $N$ input units to permutations with each input participating individually, thus indicating the weight of the contribution assigned to input unit $i$ within set $S$. Finally, the contribution value of the $i$-th input unit $\phi_{i}$ is determined.

Considering the structure and significance of formulas (2) and (4), the literature \cite{b17} proposed a shapley additive explanations(SHAP) model based on the additive feature attribution of shapley value
\begin{equation}
\phi_{i}(f,x)=\sum_{z\subseteq x'}\frac{(N-|z|-1)!|z|!}{|N|!}[g(z)-g(z\backslash\{i\})]. \label{eq}
\end{equation}
The $\phi_{i}$ value acquired via  signifies the contribution of the $i$-th input feature of the model to the ultimate classification outcome. One key advantage of this model is that, unlike other feature attribution methods, SHAP is the sole interpretation technique that fulfills the criteria of local accuracy, missingness, and consistency.
%making it theoretically more robust and consequently the most extensively utilized method. 

\textbf{Negative information:}
Negative information typically refers to the negative impact of certain features on model prediction results, and thus can be well depicted by SHAP.

The shapley value obtained from formulas (5) serves as a uniform measure of feature importance in the SHAP method. A positive shapley value for a feature indicates it has a positive role in the model's predictions, while a negative value indicates a negative role, and the presence of the feature leads to a lower prediction.
 Negative information helps identify features contributing to poor predictions or outcomes and measues the adverse impact caused by these features in the result, aiding in model improvement and decision-making.

\subsection{FGSM Attack}
DL-based AMC is vulnerable to various network attacks, which can significantly degrade their accuracy.
In training the model, the accuracy of input data is constrained. Considering the original input as $x$ and the adversarial input after disturbance $\eta$ as $\tilde{x}=x+\eta$, with $\lVert\eta\lVert_{\infty}<\epsilon$ to limit the value of $\eta$, the discrepancy between $x$ and its approximation $\tilde{x}$ can be reduced below the feature accuracy threshold and effectively disregarded. During data transmission across network layers, a dot product operation with weight vector $\omega$ is required. The resultant adversarial input post-operation is
\begin{equation}
	\omega^{T}\tilde{x}=\omega^{T}x+\omega^{T}\eta. \label{eq}
\end{equation}
The perturbation extends from the initial $\eta$ to $\omega^{T}\eta$. Consequently, introducing imperceptible minor perturbations into data can significantly impact output results due to multi-dimensional superposition, thereby degrading the model accuracy.
In multi-classification tasks, the goal is to minimize the loss function 
\begin{equation}
	\mathop{\arg\min}\limits_{\theta}\mathcal{L}(f(\theta,x),y), \label{eq}
\end{equation}
where \(f(\cdot)\) denotes the network model, $y$ is the label for input $x$, and $\theta$ represents parameters requiring optimization. Adversarial attacks amplify the loss function by introducing a slight perturbation $\eta$ to the input, resulting in the classification errors
\begin{equation}
	\mathop{\arg\max}\limits_{\theta}\mathcal{L}(f(\theta,x+\eta),y). \label{eq}
\end{equation}
The fast gradient sign method (FGSM) optimizes perturbation $\eta$ through gradient ascent during back propagation. The formula is
\begin{equation}
	\eta=\epsilon sign(\nabla_{x}\mathcal{L}(\theta,x,y)), \label{eq}
\end{equation}
where $\nabla_{x}\mathcal{L}$ is the partial derivative of the loss function $\mathcal{L}$ with respect to input $x$, $sign(\cdot)$ denotes the sign function, and $\epsilon$ is the perturbation step size. The adversarial samples $\tilde{X}$ generated by FGSM are expressed as
\begin{equation}
	 \tilde{X}=x+\eta. \label{eq}
\end{equation}

\section{Proposed SHAP-FT Defense Approach}

\subsection{Channel Transmission Model}

The transmitting source modulates the signal before transmission, and the receiver demodulates it upon reception. The signal can be defined as $s=[s[0],...,s[L-1]]$, where $L$ represents its length. The transmission channel introduces sampling rate offset, center frequency offset, selective fading, and additive white Gaussian noise to the signal. As a result, the signal received by the receiver after the $k$-th transmission can be represented as 
\begin{equation}
	x[k]=s[k]*h[k]+n[k],\label{eq}
\end{equation}
where $h$ denotes the channel response, $n$ signifies the noise with distribution $\mathcal{CN}(0, N_0)$.
% The received signal must undergo modulation and classification for more accurate decoding, known as automatic modulation classification. 
DL-based signal modulation classification offers improved performance compared to traditional methods by extracting complex features and relationships from the original signal, enhancing generalization under complex channel conditions and noise interference. 
%DL based on neural networks is considered a black-box model because it may learn incorrect relationships. This can compromise the model's ability to generalize and reduce human trust, especially after a targeted cyberattack.

\subsection{SHAP-FT Algorithm Framework}
Existing defense methods improve the network performance against attacks while have notable shortcomings. Most require substantial computational resources, such as adversarial training with the tedious training time. Many methods lack generalization and are effective only against specific attack types, while others struggle with parameter tuning and robustness.

To address these issues, this paper proposes a defense method, SHAP-FT, based on SHAP analysis and model guidance. The algorithmic framework is illustrated in Fig. 1 and is divided into three stages: Stage A generates adversarial samples to simulate realistic attacks, followed by Stage B constructing the SHAP interpreter. Finally, Stage C  fine-tunes based on the new data sequence refined by the model guided by the SHAP method and classifies the adversarial samples using the new model.
% these issues can respectively impact the model's ability to generalize and its reliability in varied contexts, while also potentially leading to a significant decline in accuracy. 
\begin{figure}[htbp]
	\centering
	\includegraphics[width=0.45\textwidth]{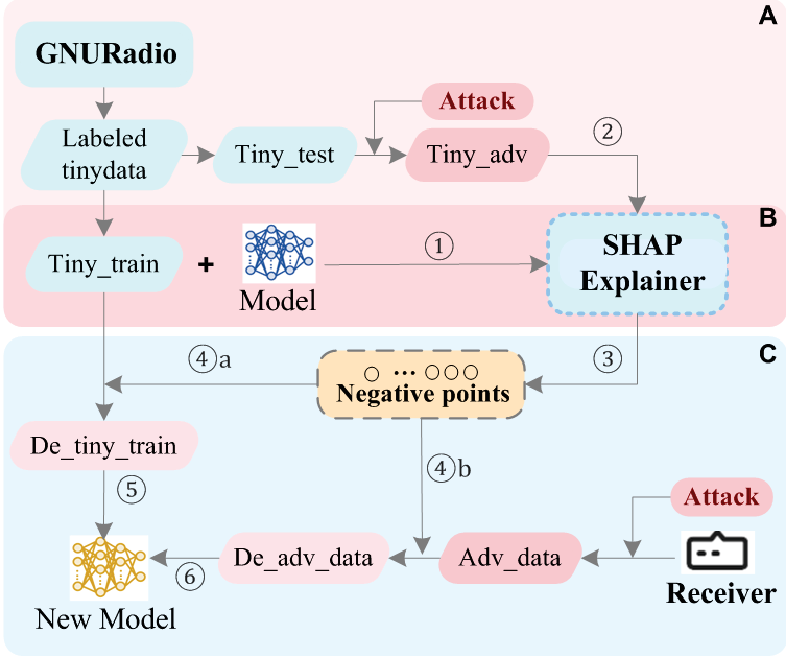}
	\caption{SHAP-FT algorithm framework.}
	\label{fig}
\end{figure}

More specifically, first, GNURadio is used to generate a tiny labeled datased, which is divided into the training set, called tiny-train, and the testing set, called tiny-test, respectively.
Tiny-train and the original attacked model are used to construct the SHAP interpreter(\textcircled{\scriptsize{1}}). Then, the tiny-test is attacked to obtain adversarial samples, tiny-adv, which are fed into the interpreter to compute shapley values for all sample features(\textcircled{\scriptsize{2}}).
Next, the sampling data with the small shapley values are regarded as the negative information reflecting the attack and misleading the classification result(\textcircled{\scriptsize{3}}). These negative data points in tiny-train are deleted, yielding a new dataset de-tiny-train(\textcircled{\scriptsize{4}}a). The original model is then fine-tuned with de-tiny-train to produce a new model(\textcircled{\scriptsize{5}}).
Finally, for the realistically attacked signal sample adv-data, the negative data points are removed to produce de-adv-data(\textcircled{\scriptsize{4}}b). These purified data are then input into the new model for classification(\textcircled{\scriptsize{6}}), enhancing the defense against the attack and improving the model performance. Additionally, the removal of some sampling data can reduce the model complexity.

\subsection{Expected Gradients}
The core calculation of the SHAP method can be conducted from different perspectives.
In the SHAP-FT, the expected gradients method\cite{b18} is used to compute the shapley values, with the gradient explainer selected as the corresponding interpreter.

The expected gradients method extends the integral gradient (IG) approach, utilizing the model's gradient information to fit the shapley values of the features. The IG method measures each feature's contribution to the prediction by integrating the gradient of the input features along the path from the baseline, which is typically set to all zeros or mean values, to the actual input. For a given model $f$ and input $x$, the IG is defined as 
\begin{equation}
IG_i(x) = (x_i - x_i') \times \int_{0}^{1} \frac{\partial f(x' + \alpha \times (x - x'))}{\partial x_i} \, d\alpha,
\end{equation}
where $x'$ is the baseline input, and $\alpha \in [0,1]$ is a scalar representing the path from the baseline to the input. The equation ultimately yields the integral of the gradient of the input $x$ as the feature contribution.

For each input sample $x$, the expected value of the IG is computed over all baseline samples.
%\begin{equation}      
%\begin{split}
%	& \text{ExpectedGradients}(x) :=  \hfill \\
%	& \int_{0}^{1} \left[ (x_i - x_i') \times \frac{\partial f(x' + \alpha \times (x - x'))}{\partial x_i} \right] d\alpha \, p_D(x) \, dx 
%\end{split}
%\end{equation}
 In practice, due to the time and memory consumption for computing the integrals, we use the sampling method to approximate these multiple integrals, which allows for the quick and accurate imputation computations calculation. That is,
\begin{equation}      
\begin{split}
& \text{ExpectedGradients}(x) \approx \mathbb{E}_{x' \sim D_{L} \cap U(0,1)}  \\
& \quad\quad\quad\quad\quad\left[ (x_i - x_i') \times \frac{\partial f(x' + \alpha \times (x - x'))}{\partial x_i} \right],
\end{split}
\end{equation}
where $p_{D}(x')$ is the distribution of the baseline.
Finally, the expected gradients of all samples are accumulated and normalized to obtain the shapley value approximation for each feature
\begin{equation}  
	\phi_i \approx \frac{1}{n} \sum_{j=1}^{n} \text{ExpectedGradients}_i(x_j). 
\end{equation} 
 
For signal data, the sampling points represent the input features of the data with $\phi_i$ denoting the shapley value of the \( i \)-th sampling point. This value measures the impact of the sampling point so that the negative information therein can be revealed on the classification result.

\subsection{Network Structure and Parameters}
Considering the temporal correlation of the signal, the convolutional and long short term memory (LSTM) are orchestrated to design the modulation classification model, as illustrated in Fig. 2.
The one-dimensional convolutional (Conv1D) layer has 128 kernels with length 8, and uses the rectified linear unit (ReLU) activation function. The subsequent LSTM layer has 128 units and is followed by a sumlayer for the dimensionality reduction. The BN layer normalizes the previous output. 
Subsequently, a fully connected layer with 256 units and the activation function Relu is connected, and finally another fully connected layer with 11 units and the activation function softmax is connected to output the result.
In the training stage, the loss function is categorical crossentropy, the optimizer is Adam with a learning rate of 0.001, the batch size is 200, and there are 200 epochs.
\begin{figure}[htbp]
	\centering
	\includegraphics[width=0.35\textwidth]{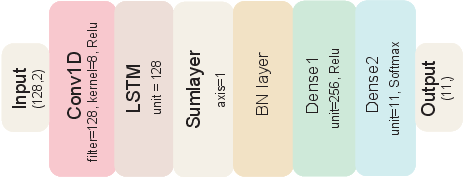}
	\caption{Network structure.}
	\label{fig}
\end{figure}

\section{Experiments and Results}
\subsection{Datasets}\label{AA}
The experiments were conducted using the RML2016.10a dataset, including 8 digital modulated signals and 3 analog modulated signals. The signal-to-noise ratios range from -20dB to 18dB in the step length of 2dB. The data format is the in-phase and quadrature (IQ) sample with a shape of $128\times2$, indicating that each sample has 128 sampling points, with the corresponding magnitude value and phase value.

The samples correspond to each modulation type and each signal-to-noise ratio are selected as the tiny data set for the experiment. The training and validation sets, collectively called tiny-train, consist of 7700 samples. Due to the limitations of the SHAP interpreter, 5000 samples of tiny-train are selected to train the gradient interpreter.
Considering that the computation time of the interpreter is proportional to the input samples, only 330 samples are selected as tiny-test. Additionally, 6600 samples are selected to simulate the real signal adv-data under attack.

\subsection{Experimental Procedure}
As shown in Fig. 1, the gradient explainer is first created with 5000 samples from tiny-train and the original modulation classification model. Then, the FGSM attack is applied to tiny-test, and the resulting adversarial samples tiny-adv are input into the interpreter for computation. The corresponding shapley values with the dimension of $330\times128\times2\times11$ are obtained. With 11 representing the predicted modulation category, as shown in Fig. 3. 
For instance, the $(i,j,k,l)$-th shapley value indicates the contribution of the angle or phase $(k=0/1)$ of the \( j \)-th sampling point from the \( i \)-th sample in the tiny-test to the prediction of that sample belonging to the \( l \)-th modulation.
 
%A positive value indicates a positive effect, a negative value indicates a negative effect, and a larger absolute value indicates a stronger positive or negative effect of the feature.
\begin{figure}[htbp]
	\centering
	\includegraphics[width=0.37\textwidth]{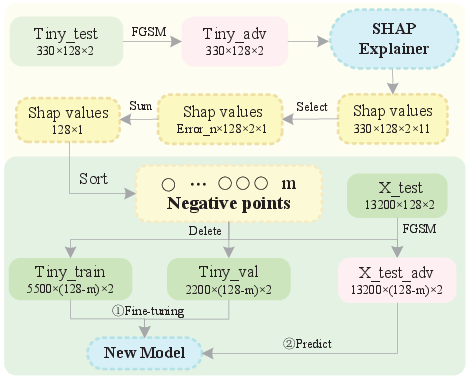}
	\caption{Experimental process.}
	\label{fig}
\end{figure}

Next, the incorrectly predicted samples in tiny-adv, denoted as error-n, were selected. These samples are dominated by the impact of the attack making it easier to identify the negative information. The shapley values of the true labels corresponding to the incorrectly predicted samples are then extracted. The feature sampling point dimension is kept while summing over the other dimensions, resulting in the sum of the shapley values of the 128 sampling points across all samples.
The $m$ sampling points with negative values indicate that the corresponding features negatively impact the classification result. These points are recorded as negative information.

The model is then fine-tuned based on the refined data excluding these $m$ negative points.
The fine-tuning were set as follows. Set the epochs to 50, change the batch size to 20, reduce the number of units in the first fully connected layer from 256 to 128, and apply early stopping. 
After fine-tuning, the new model is obtained. The adv-data is then generated by applying the FGSM attack to 6600 test samples, simulating a large number of adversarial samples received during a realistic attack. Finally, the feature sampling points corresponding to negative points are deleted from the adv-data and input into the new model for modulation classification.

\subsection{Validation of the SHAP}

\subsubsection{Negative information contained in SHAP under different attacks}
The FGSM attack with interference levels of 0.025, 0.05, 0.075, and 0.1 is applied to the tiny-test.
% to obtain tiny-adv samples under different attacks. These are inputted into the interpreter, and 
The data processing method described previously is followed to obtain the shapley summation values of 128 feature sampling points, as shown in Fig. 4.
\begin{figure}[htbp]
	\centering
	\includegraphics[width=0.4\textwidth]{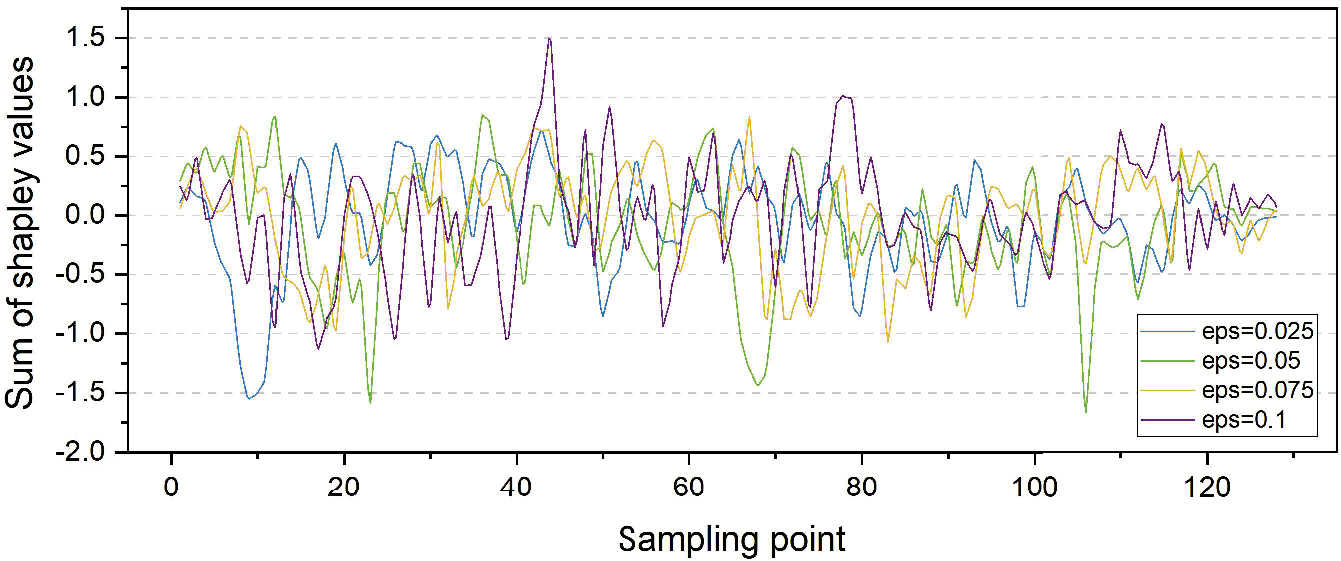}
	\caption{Sum of shapley values of 128 feature sampling points at different interference levels.}
	\label{fig}
\end{figure}

It can be observed that under different attack levels, some sampling points have negative shapley values, indicating a negative impact on the classification results. This demonstrates the effectiveness of the SHAP method in the scenario presented in this paper.

\subsubsection{Consistency of shapley value heatmap with classification results}
After computing the shapley values for tiny-adv at a attack level of 0.025, we can generate a heatmap by predicting  shapley values of all samples. This heatmap represents the aggregated shapley values for each modulation. Furthermore, we plot the confusion matrix derived from predicting tiny-adv using the original model, as illustrated in Fig. 5.
\begin{figure}[!t]
	\centering
	\subfloat[Shapley heatmap]{
		\includegraphics[scale=0.45]{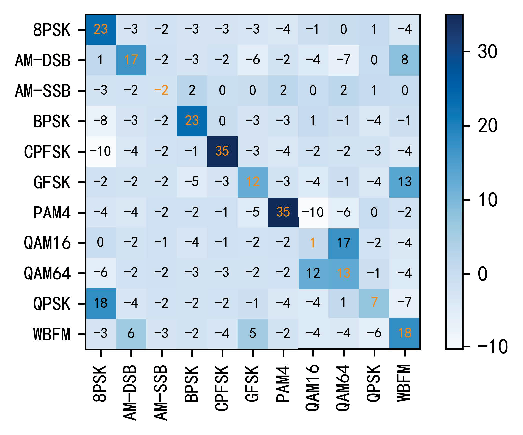}}
	\subfloat[Confusion matrix]{
		\includegraphics[scale=0.45]{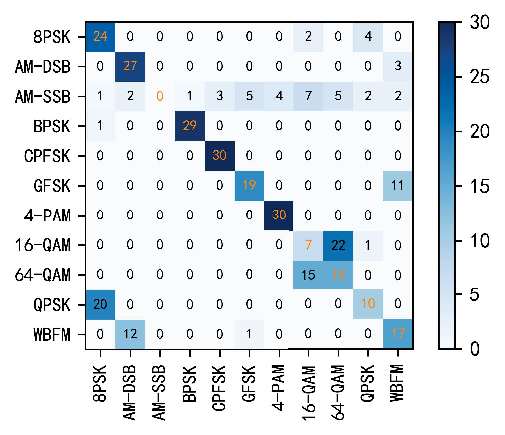}}
	\caption{The heatmap of Shapley sum value and confusion matrix of tiny-adv.}
	\label{fig}
\end{figure}

Observation reveals that the shapley value at $(i,j)$ in the left subfigure in Fig. 5 indicates the sum of the shapley values of all sampling points of all samples corresponding to the case of predicting the $i$-th modulation as the $j$-th modulation, and a higher value indicates more feature factors that play a positive role. The size of the number at $(i,j)$ in the confusion matrix on the right indicates the number of samples for which the model predicts the $i$-th modulation to be the $j$-th modulation, and a higher value indicates a higher number of corresponding predictions.

It is obvious that the distribution of the two plots is the same, which illustrates that the higher the shapley value at the $(i,j)$ position, the more positively the features play a role in classifying the $i$-th modulation as the $j$-th modulation, and thus more samples of the $i$-th label are predicted to be of the $j$-th modulation.
This demonstrates the reliability of the SHAP method.

%All correctly predicted samples in Adv-test1 are summed with 128 sample point features retained, and the frequency of the 128 sample points being disturbed in all samples is also counted. Fig.5 is obtained after normalizing the shapley summed values and disturbed frequencies of the sample points.

%\begin{figure}[htbp]
%	\centerline{\includegraphics[scale=0.3]{shap and fre.png}}
%	\caption{The sum of the shap values and the perturbation frequency of Adv-test1 at the sampling point are normalized.}
%	\label{fig}
%\end{figure}

%In general, it is logically reasonable that sampling points with lower shapley values correspond to a higher frequency of interference, and all of the above proves the instructability of the SHAP method in general.

\subsubsection{Consistency of classification results between tiny-adv and adv-data}
Due to the uncertainty of the attacked data samples received, direct SHAP analysis on realistic data is not practical. It is shown that similar data exhibit similar classification result, which enables us to use negative information from known tiny-adv samples to guide  unknown realistic adversarial samples in the adv-data.
The confusion matrix after model prediction for tiny-adv and adv-data is shown in Fig. 6.
\begin{figure}[!t]
	\centering
	\subfloat[tiny-adv]{
		\includegraphics[scale=0.45]{tinyconfusion.eps}}
	\subfloat[adv-data]{
		\includegraphics[scale=0.45]{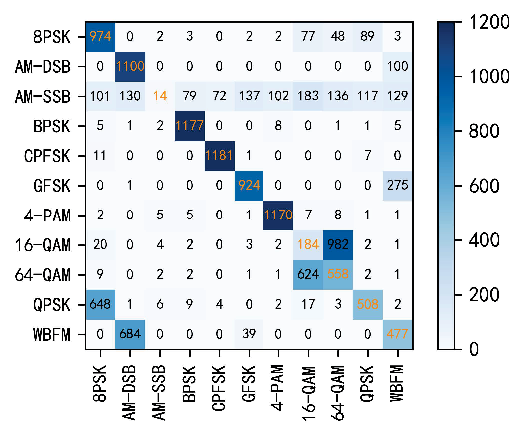}}
	\caption{The confusion matrix predicted by tiny-adv and adv-data.}
	\label{fig}
\end{figure}
It's evident that the overall distributions of the two datasets, post-attack, show similar patterns in correct and incorrect model predictions in Fig. 6. In conclusion, the negative sampling points derived from the SHAP analysis of tiny-adv can serve as valuable guidance for optimizing the adv-data.

\subsection{Experimental Results and Analysis}
Applying different interference levels of FGSM attacks and obtain the corresponding negative points, the number of negative points \( m \) under the four attack levels are 73, 78, 64, and 60, respectively. Considering the different proportions of correct and incorrect predictions among the total samples under different perturbation, it is observed that for the more severe attacks of 0.075 and 0.1, the samples have a larger proportion of incorrect predictions. Therefore, removing all the negative points can significantly improve overall performance.
For the attacks under the attack levels of 0.025 and 0.05, the relative proportion of correctly predicted samples is high. The \( m \) negative sampling points are from the incorrectly predicted samples. These points might play a positive role in the correctly predicted samples. Removing them all may have an impact on the samples that were predicted correctly.
Therefore, for the case of low attack level, all samples are selected to generate negative points. The model is fine-tuned according to the procedure in Stage C of Fig. 1. The adv-data, which simulates realistically attacked data, is processed by deleting the corresponding negative sampling points to obtain de-adv-data. Then de-adv-data is input into the fine-tuned model for predicting, and the results are shown in Fig. 7.
\begin{figure}[htbp]
	\centerline{\includegraphics[scale=0.4]{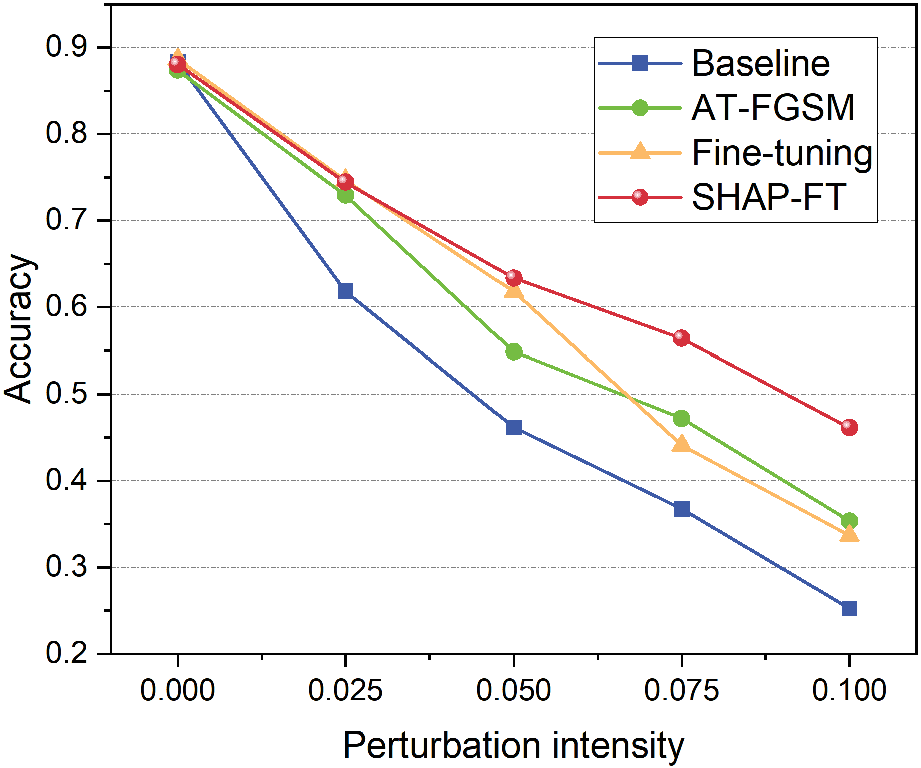}}
	\caption{Model performance under different FGSM attack levels}
	\label{fig}
\end{figure}

Fig. 7 demonstrates the classification performance under different attack levels when de-adv-data is input into the original model, adversarial training model, directly fine-tuned model, and the SHAP-FT method. 
It can be seen that direct fine-tuning performs better than the classical method AT-FGSM under low attack levels, but inferiors to AT-FGSM under high attack levels. Conversely, SHAP-FT performs close to direct fine-tuning under low attack levels, but performs significantly better under high attack levels. SHAP-FT outperforms AT-FGSM at all attack levels, proving that the SHAP-FT method offers greater defense against attacks.
Moreover, the data is compressed as some features guided by negative sampling points are removed, enhancing the model's robustness while reducing resource consumption for data storage and computation.

\section{Conclusion}
In this paper, we propose the SHAP-FT method to enhance the defense ability of AMC models against adversarial attacks. By utilizing SHAP for feature analysis, we fine-tune the model based on negative information. Experimental results demonstrate the effectiveness of SHAP-FT in improving defense capabilities while addressing challenges such as the high time complexity and parameter tuning difficulty. Additionally, SHAP-FT reduces the model complexity, making it suitable for the practical implementation.

\end{document}